\documentclass[aps,prb,showpacs,byrevtex,twocolumn,preprintnumbers]{revtex4}

\usepackage{graphicx} 
\usepackage{color}
\begin{document}

\title{The interlayer cohesive energy of graphite from \\ thermal desorption of polyaromatic hydrocarbons}
\author{Renju Zacharia$^1$, Hendrik Ulbricht$^1$ and Tobias Hertel$^{\rm 1,2}$}
\affiliation{$^1$Department of Physical Chemistry,
Fritz--Haber--Institut der Max--Planck--Gesellschaft, Faradayweg
4--6, D-14195, Berlin Germany \\ $^2$Department of Physics and
Astronomy, Vanderbilt University, Nashville, TN, USA}
\begin{abstract}
We have studied the interaction of polyaromatic hydrocarbons
(PAHs) with the basal plane of graphite using thermal desorption
spectroscopy. Desorption kinetics of benzene, naphthalene,
coronene and ovalene at sub-monolayer coverages yield activation
energies of 0.50\,eV, 0.85\,eV, 1.40\,eV and 2.1\,eV,
respectively. Benzene and naphthalene follow simple first order
desorption kinetics while coronene and ovalene exhibit fractional
order kinetics owing to the stability of 2-D adsorbate islands up
to the desorption temperature. Pre-exponential frequency factors
are found to be in the range $10^{14}$-$10^{21}\,{\rm s}^{-1}$ as
obtained from both Falconer--Madix (isothermal desorption)
analysis and Antoine's fit to vapour pressure data. The resulting
binding energy per carbon atom of the PAH is $52\pm$5 meV and can
be identified with the interlayer cohesive energy of graphite. The
resulting cleavage energy of graphite is
$61\pm5$~meV/atom which is considerably larger than previously
reported experimental values.
\end{abstract}

\date{\today}
\maketitle

\section*{I. INTRODUCTION}
The cohesive energy of a solid is determined by the interactions
among its constituent atoms. In the case of graphite, a layered
material with strongly anisotropic bonding it appears natural to
distinguish the total from the interlayer cohesive energy. The
first is dominated by strong localized covalent bonding through
$sp^2$ carbon orbitals while the latter is dominated by weak
non-local van der Waals (vdW) interactions between graphene
sheets. A description of the cohesive energy of graphite thus
necessarily involves interactions of fundamentally different
character and poses a true challenge even to the most advanced
calculational techniques. In particular dispersion forces that
give rise to the long-range attraction between graphene layers
have been notoriously difficult to predict. Not too surprisingly,
one finds that values calculated from semi-empirical or {\it ab
inito} methods for the interlayer cohesive or exfoliation energy
of graphite range from as little as 8\,meV/atom up to
170\,meV/atom.\cite{Brennan52,Allinger77,Schabel92,Trickey92,Charlier94,Rydberg03}
Experimental determinations of the interlayer cohesive energy of
graphite have been comparatively rare and are restricted to a heat
of wetting experiment by Girifalco which yields an exfoliation
energy of 43\,meV/atom \cite{Girifalco54,Girifalco56} and a
measurement by Benedict {\it et al.} based on radial deformations
of multi-wall carbon nanotubes which yields
35\,meV/atom.\cite{Benedict98} At this point it seems that not
only the agreement between theory and experiment leaves room for
improvement but that the experimental evidence for such
comparisons should also be put on a firmer basis.

\indent Here, we aim at a better experimental characterization of
the weak interlayer interactions in graphite using thermal
desorption of thin films of polyaromatic hydrocarbons from the
surface of a highly oriented pyrolytic graphite (HOPG) sample.
Polyaromatic hydrocarbons (PAHs) are planar aromatic molecules
formed by two or more fused aromatic rings and valencies of
peripheral atoms are satisfied through covalent bonded hydrogen
atoms. The interaction of PAHs with graphite is thus considered as
suitable model system for the interaction between graphene layers.
This analogy is also supported by their striking structural
resemblance, i.e. the same hybridization of atoms and practically
identical bond lengths as well as the formation of adsorbate
layers commensurate with the substrate \cite{
Gameson86,Zimmermann92,Walzer98} which suggests that the character
of the interaction between PAHs and graphite -- in particular for
larger PAHs -- should be the same as that of the interaction of a
graphene sheet with a graphite substrate. Moreover, the electronic
structure and density of states of larger PAHs converges rapidly
to that of graphene.\cite{Ruuska01}

\indent A better understanding of the long range vdW forces in
graphitic systems is also of interest when interactions between
carbon nanotubes (CNTs) are studied, for example. The latter are
frequently found to be agglomerated in so called carbon nanotube
ropes, quasi-crystalline arrangements of close packed CNTs which
are difficult so separate due to considerable long range vdW
interactions.\cite{Thess96} Only recently have soaps been
successfully used to separate and exfoliate such
ropes,\cite{OConnel02} where again, vdW forces between the
surfactant and the tubes play a crucial role as for the wider
field of soft matter physics.


\section*{II. EXPERIMENTAL}

Thermal desorption (TD) experiments were performed under
ultra-high-vacuum (UHV) conditions where the base pressure of
below $1 \times 10^{-10}$ mbar was maintained using a combination
of membrane, turbo-drag and turbo-molecular pumps. The HOPG sample
from Advanced Ceramics (Grade ZYB) was mounted on a Ta disk using
conducting silver epoxy and was freshly cleaved prior to transfer
into the vacuum chamber. The sample surface was cleaned prior to
dosing by repeated annealing cycles to 1200 K. A type-K
thermocouple was spot-welded to the Ta disk to measure and allow
control of the sample temperature. The thermocouple was calibrated
using desorption of Xe multi-layers in combination with their well
known heat of sublimation. The sample holder was attached to a He
continuous flow cryostat that enabled sample cooling down to 30 K.

\indent Benzene or naphthalene (99.89\% and 99.99\%, Aldrich)
vapour was admitted from a gas reservoir to the sample surface
through a retractable pin-hole doser.  Unwanted atmospheric
contaminants were removed from the solvents by freeze-pump cycles
prior to adsorption experiments. Exposure of HOPG to coronene
(99\%, Aldrich) and ovalene (99.5\%, Dr. Ehrenstorfer GmbH) was by
sublimation of powder material from a PID temperature controlled
Knudsen cell. A typical coverage series of TD traces was obtained
by dosing the graphite surface with successively increasing
adsorbate quantities up to a total coverage of approximately 5 ML.
TD spectra were recorded for a constant heating rate between 0.75
K/s and 2 K/s. Desorption of species with a mass to charge ratio
of up to 200 a.m.u/e, such as benzene and naphthalene, could be
monitored using a quadrupole mass spectrometer (Spectra
Satellite). Due to high molecular mass, coronene and ovalene
desorption signals had to be obtained from the total pressure
inside the UHV chamber as monitored by a Bayard--Alpert ionization
gauge. Background correction of the latter TD traces was carefully
cross-checked with simultaneously recorded TD-spectra from species
with masses between 4 to 200 a.m.u. Further details of the
experimental procedure are available elsewhere.\cite{Ulbricht02}

\section*{III. RESULTS AND DISCUSSION}

\begin{figure}[!]
\includegraphics[width=8cm]{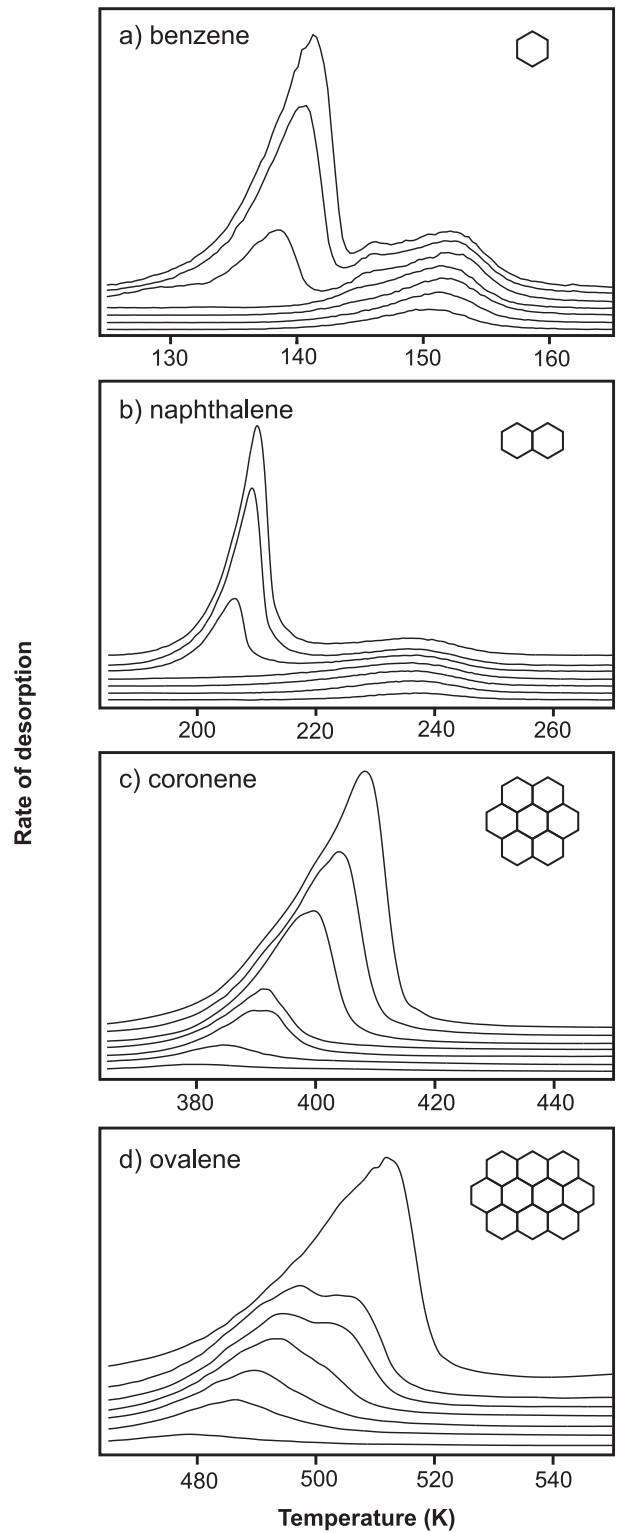}
\caption{Series of thermal desorption spectra of a) benzene, b)
naphthalene, c) coronene and d) ovalene from HOPG. The rate of
desorption was 0.7~Ks$^{-1}$, 1.0~Ks$^{-1}$, 2.0~Ks$^{-1}$ and
2.0~ Ks$^{-1}$ respectively. Coverages range from the submonolayer
regime to about 5 monolayers.}\label{pah_tds}
\end{figure}

\subsection{Thermal desorption of PAHs}

We begin our discussion with benzene TD spectra shown in
fig.\,\ref{pah_tds}a), which were recorded after exposure of the
HOPG surface with a dose of up to 5 Langmuirs (1 L = 1 $\times$
10$^{-6}$ Torr s). The TD spectra are characterized by a high
temperature feature attributed to desorption from the first
monolayer and a low temperature feature due to the desorption from
multi-layers, which are each centered around 152 K and 142 K,
respectively. The sub-monolayer phase diagram of benzene on
graphite shows a complex pattern of phase transitions were in the
high coverage- and high temperature- phases benzene molecules tend
to be oriented with their molecular plane perpendicular to the
graphitic surface to satisfy steric and entropic
constraints.\cite{Meehan80} However, at coverages significantly
below one monolayer and at 152 K, i.e. the temperature of
desorption in our experiments, the molecules are expected to be
adsorbed with their aromatic rings parallel to the
surface.\cite{Meehan80} The shape of TD spectra and the coverage
independent desorption peak maximum at low coverages are clearly
indicative of first order kinetics. With coverages approaching the
monolayer regime, however, the desorption traces broaden towards
the low temperature side and become more complex -- as expected
due to the changes of molecular orientation within the adsorbate
layer at higher coverages. The TD spectrum from a complete
monolayer is here associated with the last trace before the
additional desorption feature around 140 K develops with
increasing coverage. The latter is typical of desorption from
multi-layers with a common leading edge corresponding to zero
order kinetics.

\indent As in the case of benzene, the coverage series for
naphthalene desorption exhibits two clearly distinguishable
desorption features corresponding to mono- and multilayer
desorption respectively (see Fig.\,\ref{pah_tds}b). The TD spectra
of coronene and ovalene in Fig.\ref{pah_tds}c) and d) on the other
hand do not exhibit clearly distinguishable mono- and multilayer
desorption features. At low exposure, desorption traces from both
substances exhibit a behaviour indicative of fractional order
kinetics. Here, the saturation of the first monolayer is estimated
from the onset of multilayer desorption as apparent from the
development of a common leading edge in the TD spectra at higher
coverages. This leaves us with considerable uncertainty of the
actual monolayer coverage which -- as a worst case scenario -- is
assumed to be at most a factor of two higher or smaller. For
ovalene, further evidence for the formation of the first monolayer
is obtained from the appearance of a shoulder in the TD spectra
before the common leading edge develops. The monolayer desorption
maxima for coronene and ovalene are found at 390\,K and 490\,K,
respectively. Adsorption of coronene monolayers on graphite
reportedly also leads to the formation of 2-D islands
\cite{Walzer98} owing to lateral interaction between adsorbed
molecules, which in the case of coronene and ovalene may make
islands stable up to typical desorption temperatures. Under such
circumstances, diffusion from the island edges may become rate
limiting and give rise to fractional order desorption kinetics as
observed for coronene and ovalene (see analysis in the next
section).

\subsection{Determination of frequency factors \\ and
binding energies}

The rate of desorption of some adsorbate from a solid surface is
commonly described by an Arrhenius equation:

\begin{equation}
\frac{d\theta}{dt} = -\nu \theta^{n}\exp({-E_{a}/k_B T})
\label{WignerEqn}
\end{equation}

\noindent where, $\nu$ is the pre-exponential frequency factor,
$\theta$ is the surface coverage and $E_{a}$ is the activation
energy for desorption.

In the following we will determine the activation energies for
desorption of each adsorbate using the series of TD traces
presented above. Based on the above Arrhenius equation, a variety
of techniques, each one with unique merits, can be used to obtain
activation energies. Here, we will focus on an analysis using
Redhead's peak maximum method \cite{Redhead62} and an isothermal
analysis introduced by Falconer and Madix \cite{Falconer77}.

\subsubsection{Redhead analysis}

The Redhead equation\cite{Redhead62} relates $E_{a}$, the
temperature at the desorption peak maximum $T_{max}$ and the
heating rate, $\beta$ as follows:

\begin{equation}
E_{a}= k_B\,T_{\rm max}(\ln (\nu\,T_{\rm max}/\beta)-3.64)
\label{RedheadEqn}
\end{equation}

 The use of the latter expression is most commonly
applied to systems with first order kinetics but can be extended
to fractional- or zero order kinetics if the desorption trace used
for determination of $T_{\rm max}$ corresponds to evaporation from
a saturated monolayer.

 A crucial factor for the analysis of TD spectra using
Redhead's peak maximum method is the availability of reliable
pre-exponentials. Commonly, the latter are assumed to be on the
order of $10^{13}$-$10^{15}\,s^{-1}$ and small uncertainties of
less than an order of magnitude will not give rise to any serious
error of the activation energy. However, the assumption of
constant pre-exponentials is sometimes
problematic,\cite{Parseba01} in particular if large adsorbates
with many internal degrees of freedom are
studied.\cite{Fichthorn02} For thermal desorption of alkane chains
from graphite for example, pre-exponential factors have been
calculated using transition state theory (TST) and were found to
increase by over four orders of magnitude with chain length from
$10^{12}\,{\rm s}^{-1}$ for methane up to $10^{16}\,{\rm s}^{-1}$
for ${\rm C_{12}H_{26}}$. \cite{Fichthorn02} Qualitatively this
can be attributed to constrains on various vibrational degrees of
freedom of the molecule in the adsorbed state. If pre-exponentials
are calculated from TST one generally computes the ratio of
partition functions in the adsorbed and the transition state. If
molecular degrees of freedom are somewhat constrained or `frozen
in' in the adsorbed state this will usually reduce the
corresponding partition function and thus tends to give higher
pre-exponential factors. When studying and analyzing the thermal
desorption of small and larger polyaromatic compounds we will thus
avoid using estimated pre-exponentials and instead use
experimentally determined values.

\indent We here assume that pre-exponential frequency factors do
not depend strongly on the film thickness and that multilayer
values can thus also be used for analysis of monolayer desorption
traces. Pre-exponentials for multilayer desorption can be obtained
using the temperature dependence of the adsorbate vapour pressure.
This is done by assuming detailed balance between the rate of
adsorption and desorption between a multilayer film in equilibrium
with its gas-phase vapour. A similar approach has previously been
used by Schlichting and co-workers.\cite{Schlichting90} The
pre-exponential frequency factor can then be expressed as:

\begin{equation}
\nu =\frac{s}{\sigma\sqrt{2\pi m k_{B}T}} \ p_{0}
\label{EinsteinmodelEqn}
\end{equation}

\noindent where $s$ is the sticking coefficient, $\sigma$ the
number of adsorbates per unit area, $m$ their mass and $p_{0}$ is
the vapour pressure at infinite temperature. The latter is
obtained from the temperature-dependence of the vapour pressure
using:

\begin{equation}
p(T)= p_{0} \exp\left(-\frac{\Delta H_{s}}{k_B~T}\right)
\label{PTdependence}
\end{equation}

Here, $\Delta H_{s}$ is the heat of sublimation at the temperature
of desorption $T_{\rm max}$. However, none of the vapour pressure
curves reported in the literature extend to the range of
desorption temperatures in these
experiments.\cite{DeKruif80,Oja98,Innokuchi52} Due to the commonly
observed increase of $\Delta H_{s}$ with decreasing temperature we
thus have to extrapolate and evaluate the vapour pressure curves
at temperatures where desorption occurs in our experiments. This
can be done by a fit to vapour pressure data using Antoine's
expression: \cite{Antoine1888}

\begin{equation}
\ln (p(T))= A-\frac{B}{T+C} \label{AntoineEqn}
\end{equation}

\noindent where A, B and C are the fit coefficients listed in
table \ref{VaporPressure}.

\begin{table}[exactlyhere]
\begin{ruledtabular}
\caption{Vapor pressures and desorption temperatures used to
compute the frequency factor}
\begin{tabular}{l c c c c c c}
Molecule    &$A$ & $B$ (K) &$C$ (K)  &
$\sigma$(cm$^{-2}$)\footnotemark[1] & $p_{0}$(mbar) & $T_{\rm
max}$(K)
\\ \colrule
Benzene       &26 &7640 &30 & 2.7$\times$10$^{18}$ & 5.8$\times$10$^{11}$         & 151   \\
Naphthalene & 43& 20100&124   & 1.6$\times$10$^{18}$  & 1.3$\times$10$^{14}$         & 235 \\
Coronene     &37 &30400& 184 & 0.9$\times$10$^{18}$  & 2.6$\times$10$^{13}$         & 390\\
Ovalene      &&&  & 0.6$\times$10$^{18}$  & 7.6$\times$10$^{13}$         & 490 \\
\end{tabular}
\label{VaporPressure} \footnotetext[1] {See the text for
references}
\end{ruledtabular}
\end{table}

\indent For ovalene, where no vapour pressure data is available,
the pre-exponential factor $5.6\times10^{21}\,{\rm s}^{-1}$ was
obtained using the tabulated slope and offset from the Clausius
Clapeyron equation in the form $\ln(p)=A-B/T$.\cite{Innokuchi52}
The density $\sigma$ of benzene, naphthalene and coronene
molecules adsorbed on the graphite surface can be obtained from
LEED or STM data.\cite{Bardi87,Walzer98} For ovalene where again
no such data is available $\sigma$ can be estimated assuming that
molecules are close packed with the polyaromatic rings oriented
parallel to the surface. The resulting densities are summarized in
table \ref{VaporPressure}.

The sticking coefficient $s$ used in eq. (\ref{EinsteinmodelEqn})
is here assumed to be close to unity which is commonly observed
for homoepitaxial growth and should also be appropriate for
adsorption of weakly interacting polyaromatics on graphite. Table
\ref{VaporPressure} summarizes all parameters used for the
computation of frequency factors as well as the temperature at the
desorption peak maximum used for computation of the activation
energy from the Redhead equation. The computed frequency factors
are summarized in table \ref{ActivationEnergy} together with the
resulting activation energies as obtained from eq.
\ref{RedheadEqn}.

\subsubsection{Falconer--Madix analysis}

\begin{figure}[!]
\includegraphics[width=8cm]{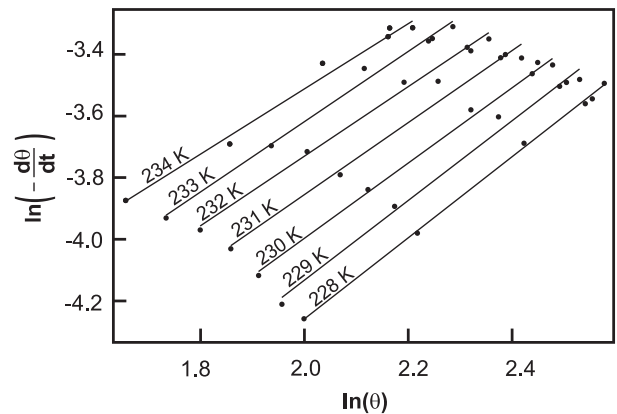}
\caption{Linear fit to isothermal desorption data for naphthalene
in the temperature range of 228 to 234 K.}\label{IsoDesoPlot}
\end{figure}

In this section we will perform an isothermal analysis of TD
spectra of the type frequently referred to as Falconer--Madix
method.\cite{Falconer77}  For this purpose we plot the logarithm
of the desorption rate $\ln(-d\theta/dt)$ versus $\ln \theta $ as
evaluated at one specific temperature for several desorption
traces of different initial coverage. A linear fit to the
resulting data sets from sub-monolayer coverages as shown in
Fig.\,\ref{IsoDesoPlot} can be analyzed according to:

\begin{equation}
\ln (-\frac{d\theta}{dt})= \ln \nu + n \ln
\theta-\frac{E_{a}}{k_B\,T} \label{FMEqn}
\end{equation}

\begin{figure}[!]
\includegraphics[width=8cm]{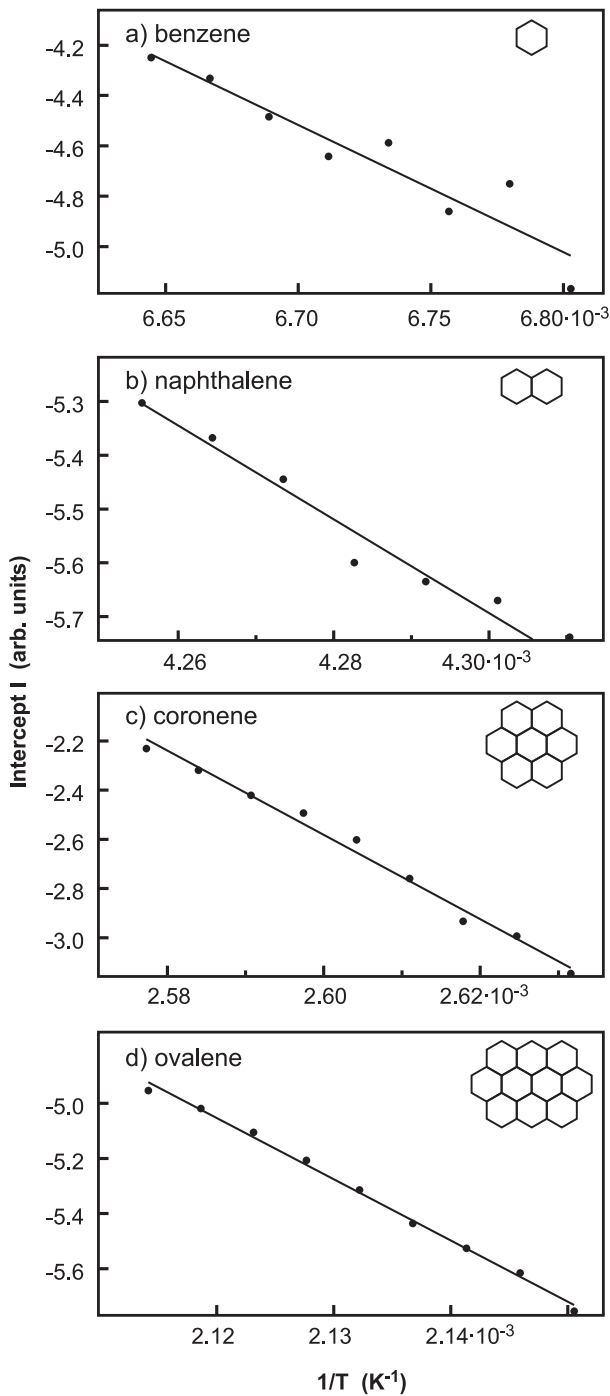}
\caption{Plot of intercept I vs inverse temperature.}
\label{InterceptPlot}
\end{figure}

\noindent If the intercept, $I$

\begin{equation}
I = \ln(\nu) -\frac{E_{a}}{k_B\,T} \label{FMInterceptEqn}
\end{equation}

\noindent is plotted versus $1/T$ (see Fig.\,\ref{InterceptPlot}
for naphthalene) one obtains the pre-exponential frequency factor
as well as the activation energy from a straight line fit to the
data. From the slope of naphthalene and benzene desorption
isotherms we obtain the order of desorption $n$ of 0.95$\pm$0.02
and $n$=1.01$\pm$0.02 respectively. On the other hand, coronene
and ovalene appear to follow fractional order kinetics with
$n$=0.27$\pm$0.04 and 0.34$\pm$0.01 respectively.

\indent The pre-exponentials as well as the activation energies
obtained from this analysis are summarized together with the
results from the Redhead peak-maximum analysis in Table
\ref{ActivationEnergy}.

\begin{table}[!]
\begin{ruledtabular}
\caption{Binding energy and frequency factor : comparison between
Redhead and Falconer--Madix analysis}
\begin{tabular}{l c c c c c}
&\multicolumn{2}{c}{Redhead} &\multicolumn{2}{c}{Falconer--Madix}
\\
&$\nu(s^{-1}$) &$E_{a}$(meV)
                                                             &$\nu(s^{-1}$) &$E_{a}$(meV)
                                                            \\
                                                 \colrule
Benzene        &1$\times$10$^{16\pm3}$      &0.50$\pm0.08$  &5.0$\times$10$^{15\pm2}$     &0.50$\pm0.08$   \\
Naphthalene    &5$\times$10$^{16\pm2}$      &0.8$\pm0.1$  &1.0$\times$10$^{17\pm1.5}$     &0.90$\pm0.07$   \\
Coronene       &2$\times$10$^{16\pm2}$      &1.3$\pm0.2$  &1.7$\times$10$^{18\pm0.5}$     &1.5$\pm0.1$  \\
Ovalene        &5$\times$10$^{21\pm3}$      &2.2$\pm0.2$  &8.0$\times$10$^{18\pm0.5}$     &1.97$\pm0.08$   \\
\end{tabular}
\label{ActivationEnergy}
\end{ruledtabular}
\end{table}

Note, that the pre-exponential frequency factors in the present
study are considerably larger than those commonly used for
desorption of smaller molecules where typical values are on the
order of $10^{13}\,{\rm s}^{-1}$ to $10^{15}\,{\rm s}^{-1}$. This
reinforces the necessity to determine these parameters accurately
and reliably either from experiments or by complementary
theoretical investigations as recently reported by Fichthorn et
al.\cite{Fichthorn02}  The high pre-exponentials found here can
qualitatively be accounted for by the larger difference between
partition functions in the adsorbed and the transition state, if
compared with desorption of mono- or diatomic adsorbates, for
example. A comparison of the pre-exponentials obtained here by the
two different methods shows good agreement within estimated error
bars.

\subsection{The interlayer cohesive
energy of graphite}

The cohesive energy of a solid is commonly referred to as the
energy needed to "disassemble it into its constituent parts"
\cite{Ashcroft} while the work of cohesion is occasionally also
referred to as the energy needed to "separate unit areas ..." of a
medium "... from contact to infinity in
vacuum".\cite{Isrealachvilli92} The two are not identical and for
a layered system with extremely anisotropic bonding like graphite
the first can be identified with the exfoliation energy $E_{ex}$,
i.e. the energy needed to separate all layers of the crystal to
infinity while the latter is equivalent to the cleavage energy
$E_{cl}$ which is commonly slightly larger than the exfoliation
energy. Previous work suggests that the energy needed to separate
a single sheet of graphene from a graphite crystal, i.e. the
exfoliation energy, is approximately 18\% smaller than the
cleavage energy.\cite{Girifalco56} The normalization in this paper
will be with respect to  surface atoms and not to area. The area
per surface atom in a graphene sheet is $((2.46\cdot 10^{-10}~{\rm
m})^2\cdot \sqrt{3})=1.05\cdot 10^{-19}{\rm m}^{2}$.

\indent In the following we will use the results from the previous
section to determine the interlayer cohesive energy of graphite to
which the dominant contribution is generally believed to arise
from long range van der Waals (vdW) interactions between graphene
sheets. Our approach for the determination of the interlayer
cohesive energy  -- using the activation energies for desorption
of small polyaromatic molecules, i.e. small `flakes of graphene'
-- is based on the near-additivity of such vdW
interactions.\cite{Isrealachvilli92,Lii89} The latter is well
established, for example, for the cohesive energy of alkanes where
deviations from linearity in the number of chain segments are
about 1 \% or less.\cite{Isrealachvilli92} The activation energy
for desorption -- as measured by TD spectroscopy -- will here be
identified with the binding energy of the adsorbate to the
graphite surface. The contribution of individual carbon atoms to
this binding energy is derived from our data which -- in the limit
of infinitely large PAHs -- would correspond to the energy needed
to separate a graphene sheet from its parent crystal and is thus
associated with the interlayer cohesive energy. Additional
contributions to the desorption energy from intermolecular
interactions are assumed to be negligible. This is justified if
adsorbate-adsorbate interactions are either small due to large
mutual separation -- as in the case of the low coverage regime for
benzene or naphthalene desorption -- or this may be justified if
adsorbate-adsorbate interactions within 2-D islands on the surface
are comparatively small as in the case of adsorbed coronene or
ovalene.

\begin{figure}[!]
\includegraphics[width=8cm]{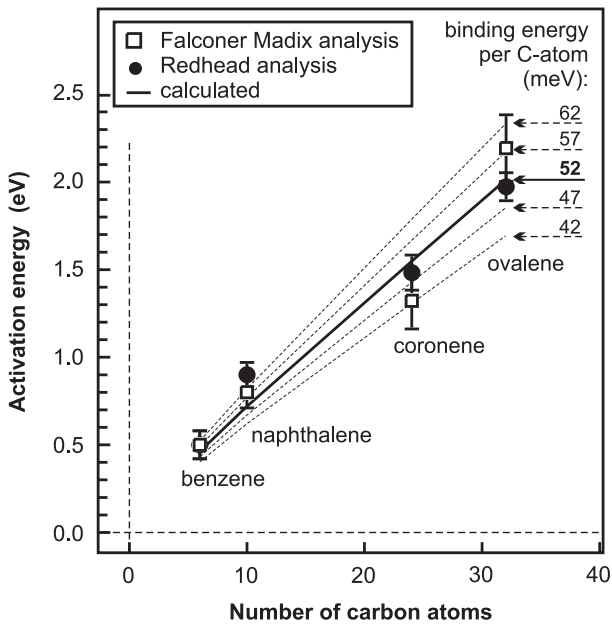}
\caption{Dependence of activation energy for desorption on the
number of carbon atoms of four polyaromatic
hydrocarbons.}\label{EaVsNumberofCarbons}
\end{figure}

\indent A rough estimate of the exfoliation energy can be obtained
by averaging over desorption energies per carbon atom for all
studied adsorbates which would yield 67~meV/atom. In this case
however, one neglects the small but significant contribution to
the binding energy from hydrogen atoms and this approach is thus
expected to overestimate the contribution of carbon atoms to the
total binding energy. A more thorough analysis of the binding
energies plotted in Fig.\,\ref{EaVsNumberofCarbons} thus also has
to account for the contribution of hydrogen atoms. Here, this is
done by optimizing the carbon-carbon and hydrogen-carbon
interaction potentials to give optimum agreement of experimental
energies with calculated binding energies. Experimental data for
this optimization are averaged over values obtained from both the
peak maximum and isothermal analysis. The calculated binding
energies are obtained in the usual manner by summation over
empirical vdW pair-potentials:

\begin{equation}
E_B^{\rm PAH} = \sum_{n,m}V^{\rm CC}(|r_n-r_m|)+ \sum_{l,m}V^{\rm
HC}(|r_l-r_m|) \label{pairpotentials}
\end{equation}

\noindent where the summation over $n$ and $l$ includes all carbon
or hydrogen atoms in the adsorbate and the summation over $m$
includes all atoms in the substrate. The carbon-carbon (C-C) and
carbon-hydrogen (C-H) vdW potentials are those of the MM3 force
field by Allinger and co-workers. \cite{Allinger89} They have the
form:

\begin{equation}
V^{\rm AB}(r)=\epsilon_{\rm AB}\left[184000\exp\left(-\frac{12
r}{r_{\rm AB}}\right)-2.25\left(\frac{r_{\rm
AB}}{r}\right)^6\right] \label{AllingerVDW}
\end{equation}

\noindent where for dissimilar atoms A and B $\epsilon_{\rm AB}$
is given by $\epsilon_{\rm AB}={\sqrt {\epsilon_{\rm A}
\epsilon_{\rm B}}}$ and $r_{\rm AB}$ is given by $r_{\rm
AB}=r_{\rm A}+r_{\rm B}$. The values ($\epsilon_{\rm C}$,$r_{\rm
C}$) and ($\epsilon_{\rm H}$,$r_{\rm H}$) given by Allinger and
co-workers for carbon and hydrogen are ($2.44\,{\rm
meV}$,$1.96\,{\rm \AA}$) and ($0.87\,{\rm meV}$,$1.67\,{\rm
\AA}$), respectively. In the following we use the depth of the two
vdW potentials $V^{\rm CC}_0$ and $V^{\rm HC}_0$ as free scalable
parameters while the position of the potential minima is kept
fixed. The binding energy $E_B^{\rm PAH}(V^{\rm CC}_0, V^{\rm
HC}_0,z)$ is then optimized by adjusting the molecule-surface
distance $z$ while the orientation of the aromatic rings of the
molecules is fixed parallel to the graphite surface. We then
determine the set of parameters $V^{\rm CC}_0$ and $V^{\rm HC}_0$
which gives best simultaneous agreement of all calculated with all
experimental binding energies by minimizing the corresponding root
mean square deviation. Using the MM3 force field parameters the
depth of the hydrogen vdW potential for interaction with a
graphite surface is estimated to be 27\,meV/atom. If we allow this
to vary by at most $\pm 5\,$meV we get best agreement with the
experimental binding energies if the depth of the carbon-graphite
potential is $(52\pm5)\,$meV/atom. From
Fig.\,\ref{EaVsNumberofCarbons}, where calculated and experimental
binding energies are compared one finds that deviations between
experimental and calculated values are about 10\% or less. The
major uncertainty here arises due to the large error bars of the
pre-exponential in the Redhead method and from the estimated error
of the temperature calibration as well as the reproducibility of
TD traces due to small variations of the sensed temperature for
the Falconer-Madix analysis. For coronene and ovalene an
additional error of about 2\% arises due to uncertainties of the
coverage calibration. The cleavage energy of graphite is obtained
from the above value by accounting for the 18\% higher energy
previously observed for the separation of two crystal halves if
compared with the separation of a single graphene layer from its
parent crystal.\cite{Girifalco56} This yields a cleavage energy of
$(61\pm 5)$\,meV/atom which is significantly larger than
previously reported values.

\indent The earliest measurement of the exfoliation energy from
heat of wetting experiments by Girifalco and Lad gave $(260\pm
30)\,{\rm ergs/cm}^2$ which -- using the carbon atom density
within a graphene sheet -- is equivalent to $(43\pm5)$\,meV per
surface atom.\cite{Girifalco56} Unfortunately, the only
documentation for the heat of wetting data used for the
determination of $E_{ex}$ by Girifalco and Lad has been published
in a thesis and is not readily available.
\cite{Girifalco54,Girifalco56} Also, no information on the kind of
graphite sample used in the original experiments is available.
Carbon powders commonly used for such experiments usually have
higher surface areas of the order of tens to hundreds of square
meters per gram. They are usually treated with acids to oxidize
non-graphitic contaminants and are then thermally treated to
achieve the highest possible degree of graphitization. The
adsorptive properties and surface chemistry of such carbons,
however, depend strongly on the history of the sample
treatment.\cite{Schlogl02} An assessment of systematic and sample
dependent uncertainties of the value reported by Girifalco is thus
difficult.

\indent An even smaller value of ($35^{+15}_{-10})$\,meV/atom for
the cohesive energy of graphite was obtained by Benedict et al.
from the analysis of collapsed multi-wall carbon
nanotubes.\cite{Benedict98} This analysis is based on a
measurement of the diameter of hollow 'bulbs' at the sides of
three different collapsed multi-wall carbon nanotubes with a
precision of about 1-2~\AA. N, other than statistical errors or
those due to the limited accuracy of the bulb diameter
determination may as well contribute to uncertainties associated
with this value. This again makes it difficult to assess the
relevance of possible systematic or statistical errors which could
help to better understand the discrepancy between our and other
experimental determinations of the cohesive energy of graphite.
Note, that a distinct advantage of the present investigation is
that experimental conditions and assumptions leading to the
conclusions are most clearly defined and that a well characterized
model system is used for the investigation of the interlayer
cohesive energy of graphite.

\indent As stated in the introduction, theoretical estimates using
ab initio or semi-empirical methods show a much stronger variation
from values as little as 8\,meV/atom to as much as
170\,meV/atom.\cite{Brennan52,Allinger77,Schabel92,Trickey92,Charlier94,Rydberg03}
These large discrepancies are partly due to the inherent
difficulties encountered in the calculation of long-range
dispersion forces. Even advanced calculations using non-local
density functional theory which account for vdW interactions with
reported values for a single pair of graphene sheets of only
$35$\,meV/atom \cite{Rydberg03} tend to underestimate the
interlayer cohesive energies. However, the desorption energies
reported here for different PAHs may serve as a useful benchmark
for future studies to allow a better comparison of theoretical
binding energies with experiment. Reliable values for the cleavage
and interlayer cohesive energies may eventually be derived from a
successful calculation of the interaction of PAHs with graphite if
the interlayer forces in graphite can be treated on the same
footing.

\section*{IV. Summary}

In conclusion, we have presented a thermal desorption study of
benzene, naphthalene, coronene and ovalene adsorbed the basal
plane of graphite. Binding energies were obtained by the
peak-maximum method and alternatively by an isothermal analysis
which also allowed to determine pre-exponential frequency factors.
For the peak maximum method we derived pre-exponentials from vapor
pressure data in combination with Antoines law for extrapolation
of the data to the temperature of desorption. Both methods
indicated that pre-exponential factors increase with adsorbate
size from $10^{14}\,{\rm s}^{-1}$ to $10^{21}\,{\rm s}^{-1}$. The
corresponding binding energies increase from 0.50~eV for benzene
to 2.1~eV for ovalene. These values were used to determine the
interlayer contribution to the cohesive energy of graphite by
assuming pairwise additivity of the interaction of carbon and
hydrogen atoms to the total binding energy of the molecules. The
resulting cleavage energy of graphite of $(61\pm 5)$\,meV/atom is
derived from the average carbon atom contribution to the binding
energy of the PAHs of $(52\pm 5)$\,meV/atom which can also be
associated with the exfoliation energy. This is significantly
larger than previous experimental determinations and these results
provide an experimental benchmark for future theoretical
investigations of interlayer bonding and van der Waals
interactions in graphitic systems.

\acknowledgments It is our pleasure to acknowledge continuing
support by G. Ertl.

\end{document}